# Nano Explosions

Sascha Vongehr,*

College of Engineering and Applied Sciences, Nanjing University, Jiangsu, P. R. China

**Abstract:** At the nano-scale, surface phenomena such as attractive VdW forces strongly dominate; explosions may well be thought impossible. We confirm nano explosions that are important for a fractal (hierarchical, scale invariant) pore structure, greatly increasing pore access. Analyzing only microscope images, image analysis and statistical error reduction algorithms alone provide conclusive evidence of explosively expelled material. The results reconfirm that computer image recognition and statistical analysis are a widely applicable and inexpensive technique for determining parameters which are otherwise unavailable, such as the densities of single nano shapes. The explosion mechanism explains optimum calcination temperatures for cobalt-hydroxide and why slow heating optimizes porosity. Slow heating increases the accessible surface area by 60% with this supercapacitor material.

## Introduction

Already in the micrometer range, VdW forces such as surface tension dominate over inertia; small aquatic animals must constantly propel themselves or else immediately come to a stop as if in thick honey rather than water. At nano sizes , it is questionable whether true explosions of material by itself (excluding high impact projectiles, strong laser beams and suchlike) occur at all. Even with very weakly bound alkali clusters on helium droplets, suggested explosively ejected regions upon internal reactions[1] have later been rejected.[2,3] Moreover, there is no easy method for establishing densities of single nano objects in a synthesis batch. The floating/sinking transition in fluids of different densities fails with porous particles. Nano-lever bending[4,5] is prohibitively expensive if desiring size-distribution statistics. With usually $10^{12}$ particles per mg, one cannot repeatedly dilute until a micro drop's particles may be countable in a Scanning Electron Microscope (SEM), because the particles stick to container walls. Washing away of un-reacted chemicals hinders comparing with calculated yields from reactant concentrations, and moisture as well as tiny amounts per sample renders weightings inaccurate. As was first shown by proving the low density of carbon in Ag nanoparticle doped carbon spheres:[6] With complex nano structures, quantitative determination of simple parameters such as a density needs analyzing features that are independent of absolute amounts. Accurate structure analysis from images and good statistics before and after reactions can provide such quantities.[7] Such parameters are very useful; the density of the nano plates discussed here already allowed proving that the linear decline of capacitance (Region III of Fig. 1a) is mainly an effect of the plates changing their shapes.[8]

---

* Corresponding authors' emails: vongehr8@yahoo.com

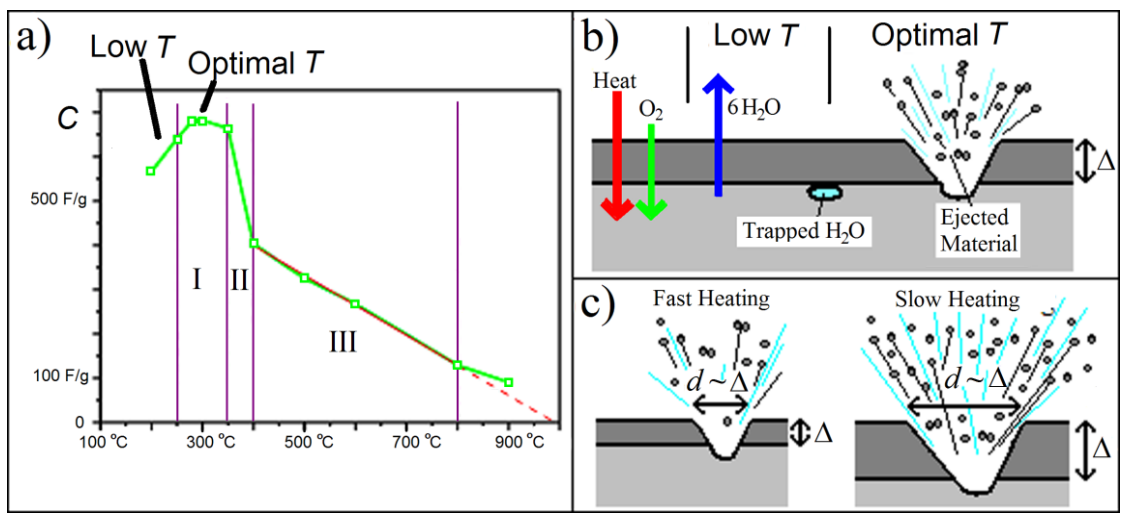

**Fig. 1:** (a) Specific capacitance $C$ of $Co_3O_4$ nano plates versus calcination temperature $T$, all heated at 10 °C/min. A steep decline (II) after the maximally porous region (I) is due to melting of pore-access (III is a shape-effect dominated decline);[8] (b) Explosion mechanism at the dense surface layer (dark grey, thickness $\Delta$) of a heated nano plate. (c) Explosion crater diameters $d$ are comparable to $\Delta$. Slow heating leads to later but larger explosions.

We will now give a new example for how integrated image analysis and statistics[7] obtains otherwise unavailable parameters. Moreover, since the accessibility of nano-pores inside larger pores is central, not even TEM images, which are two-dimensional (2D) projections of 3D structures, can supply more direct evidence, let alone values that are as precise as those from the here applied method. We simultaneously report a novel nano explosion mechanism (Fig. 1b) and its exploitation (Fig. 1c) for increasing the accessible surface of an important material for supercapacitors and lithium storage.

Soft templates[9] can obtain porous oxides such as $Al_2O_3$.[10,11] Hard templates increase stability, for instance with carbons[12,9] and cobalt oxide.[13] However, templates generally supply particular pore-sizes. It is known that porosity can also originate from explosions, for example ruptures due to trapped carbon dioxide.[14] While random growth processes result in peaked log-normal size distributions,[15,16,17] explosive fractionation leads to more scale invariant gamma distributions,[18] or even entirely peak-less power laws[19,20] and exponentials.[21,22] Such fractal structures increase the access to pores, because molecules may exit a nano pore through a succession of increasingly larger pores instead of having to pass many nano channels before reaching the exterior.

The optimum accessible pore area in $Co(OH)_2$ calcinations is obtained at 300°C.[23] Capacitance measurements (Fig. 1a) of the resulting $Co_3O_4$ concluded that higher temperatures "melt" the pores close.[8] A full understanding of the optimum temperature however also needs to explain the lack of access at temperatures that are colder than the optimum. The calcination reaction suggests involvement of explosions. We will discuss how the explosion mechanism implies that heating slower than the usual 10 to 20 °C per minute will increase the force of the explosions. This improves pore access in several ways. Measurements confirm a 60% higher accessible surface area with this important material. $Co_3O_4$ is one of the most intriguing magnetic p-type semiconductors and widely applied in lithium-ion batteries,[24] heterogeneous catalysis,[25,26] electrochemical[27] and electrochromic[28,29] devices, solid state sensors[30,31,32] and more.[33] Needles,[34] rods,[35] fibers,[36] tubes,[37] columns[38] and other shapes have been obtained through thermal conversion of shaped precursors, usually cobalt carbonates[39] or beta-cobalt hydroxide.[38] Our results should be relevant for many of such research projects.

Computer image recognition assisted shape analysis (Fig. 2) confirms the loss of explosively expelled, atomized material. Exact shapes, good statistics from many shapes, and taking the size distributions' behavior into account are all vital for this analysis. For example, it cannot be neglected that the shapes are truncated triangles instead of regular hexagons. Exploiting the information hidden in size distributions is interesting across nanotechnology. There is no comparatively simple method that allows similarly precise quantitative statements about the density of nano shapes in a batch.

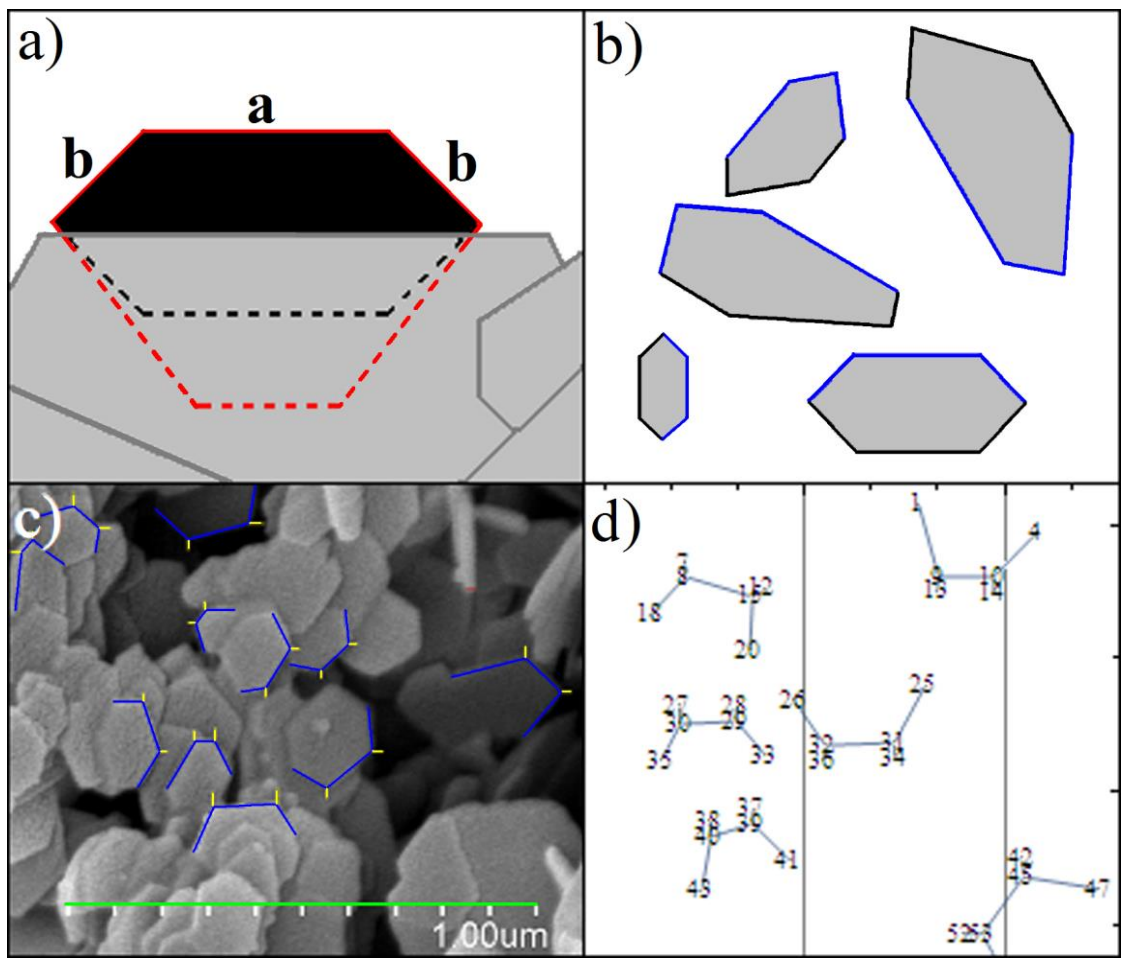

**Fig. 2:** Adapted from ref[40] with permission. (a) A shape (black) is hidden below others (grey). It could be a truncated triangle resting flat on the image-plane (red hatched line), with *a* truly longer than the adjacent sides *b*, or instead, a regular hexagon (black hatched line) rotated backward, so that *b* merely seems shorter. Distinguishing these cases necessitates two lengths and two Euler angles. (b) Testing with simulated images initially failed! One must include three adjacent sides (blue) for five degrees of freedom. (c) Part of marked SEM, showing cobalt oxide plates, ready for the computer analysis. (d) Graph analysis provides the information for vector calculations, making explicit calculations of angles unnecessary.

## The Explosion Mechanism

If cobalt-hydroxide is thermally dehydrated, the denser $Co_3O_4$ should first form in the surface because heat and oxygen enter from the outside according to[12] $6Co(OH)_2 + O_2 \rightarrow 2Co_3O_4 + 6H_2O$. This produces six water vapor molecules in exchange for every $O_2$ molecule, and additionally inside a shrinking, increasingly denser material. Gaseous water is trapped and then explodes through the dense surface layer. If this hypothesis (**H1**) is correct, material will be lost from the explosion craters. Moreover, the porosity should depend on the speed of the heating, which we predict as follows (**H2**): The surface that has already turned into the denser oxide has a certain thickness, labeled "$\Delta$", at the time when water starts to be trapped and explode through this layer (Fig. 1b). The explosion craters will be of a comparable size $\Delta$ (very roughly ~ 1 nm). Heating faster will increase the pressure below a still thinner surface layer rapidly. Therefore, $\Delta_{(fast)}$ is smaller (Fig. 1c). When heating *slowly* however, the pressure necessary to rupture the surface may not be attained until much later. The water now released at the shallow depth of the previously considered $\Delta_{(fast)}$ has time to diffuse through the surface. The dehydration goes on much further before explosions occur, hence $\Delta_{(slow)} > \Delta_{(fast)}$. Explosive rupturing of the thicker surface layer will now require a higher pressure. The thicker $\Delta$ becomes, the more difficult it is for water to escape. Rupturing will eventually occur, because temperature $T$ increases far beyond the dehydration onset at 180°C. The explosion crater diameters are now comparable to $\Delta_{(slow)}$. Slow heating therefore leads to better access to both, the interior (1) and to the surfaces between stacked shapes (2). Explosions lead generally to scale invariant structures. The surface density of small pores is therefore similar, but they distribute now in a rougher and thus larger surface, so the number of accessible small pores is increased (3). These are three ways in which better access to pores is expected. Meso-pores (2 to 50 nm diameter) are important for catalysis and supercapacity, but only if they can be accessed.

## Experimental Confirmation

The precursors are β-Co(OH)$_2$ plates from a hydrothermal synthesis described elsewhere,[8] and they are not porous. Calcination at 300°C preserves the hexagonal shapes. Shape and size analysis is based on a novel computer assisted image analysis that was custom modified for these shapes. It is able to determine geometries of single shapes accurately in spite of random rotations with unknown biases due to their alignment on top of the microscopy substrate and mutual stacking (Fig. 2). The algorithms keep track of the size distribution statistics; for details see the Supplemental Information (SI). The precursors are truncated triangles with a long edge $a$ = (130 ± 35) nm. The long to short edge ratio is $a/b$ = 1.4 ± 0.3. The assumption of regular hexagons ($b = a$) results in large, erroneous tilt angles even for shapes that are flat in the image plane, and such leads to overestimating their areas by on average 40%. Therefore, the assumption of hexagons in the literature is in this precise sense strictly wrong and would not allow our results. The up to 25% statistical variations are not so much errors but represent the physical widths of the size distributions. They are crucial data, as shall now be explained.

Whenever possible, such as in case of the face area $A = \sqrt{3}\left(a^2 + 4ab + b^2\right)/4$, the value $A$ is first calculated for every shape, *before* averaging to $\langle A \rangle$. We call this "delayed averaging". Calculating instead from the averages $\langle a \rangle$ and $\langle a/b \rangle$, the result "$A_{(<a>)}$" is only 88% of the true $\langle A \rangle$, yet again an error over 20%. The appropriate assumption for random growth processes is a log-normal (LN) size distribution. Therefore, true averages can also be calculated from systematic underestimations such as $A_{(<a>)}$ by applying[41] $\langle A \rangle_{LN} = A_{<a>} \exp\left[d(d-1)s^2/2\right]$, where $d$ is the ratio between the involved dimensions of the input and output parameters (here $d = \dim_{Area}/\dim_{Length} = 2$). Let us call this "LN-correction." The standard deviation of the logarithms of the lengths, i.e. $\ln[a]$, is here $s$ = 0.16. $\langle A \rangle_{LN}$ is therefore 98% of $\langle A \rangle$, and there is generally always only a 1 to 3% mismatch between the result from the LN-correction and the delayed averaging. The large standard 'errors,' such as in our result $A = (35 \pm 14)\ 10^3\text{nm}^2$, do not imply that the applied methods are imprecise. On the contrary, they report the actual, physical distribution widths. The distribution of the height $h$ is also broad. Therefore, the volume $\langle A \rangle \langle h \rangle$ suffers a similar systematic error. The volume cannot be calculated separately for every shape, because the heights $h$ are not extracted from the same shapes as $a$ and $b$ are. We remove the error in $h$ by LN-correction.

Chemical dehydration starts at about 180°C. The loss of weight is therefore always 13.6%, corresponding to $3\text{Co(OH)}_2 \rightarrow \text{Co}_3\text{O}_4$, regardless the heating rate, as long as the final temperature is above 250°C and one starts with dry material. Staying at 300°C guaranties that any improvement due to a different heating rate $\tau = dT/dt$ will improve the material over previous reports, but one should note that a proper 'walk-in optimization' onto a global maximum would likely result in yet better products, perhaps at 280°C.

The slow heating hypothesis H2 is easily confirmed by Brunauer-Emmett-Teller (BET) measurements (unpublished). 20 °C/min lead to the reported maximum[42] of 76 m$^2$/g at 300°C. BET measurements are usually imprecise. For example, at 10 °C/min, 98 and 113 m$^2$/g can result. However, the slow 1°C/min provides 123 m$^2$/g, and this is an improvement by two standard deviations. It will now be rigorously proven that explosions are indeed responsible. There should be atomized material lost, which is washed away when cleaning samples. Non-porous β-Co(OH)$_2$ has a density of $\rho_0 = 3.597$ g/cm$^3$. The average precursor's volume contains thus $3.7*10^{-17}$ mol of cobalt.

<u>Calcination at 200 to 250 °C</u> shrinks $a$ to $(117 \pm 30)$ nm. These values (average and width) are for example based on 109 individual shapes from five different SEM images. Height $h$ shrinks to $(25 \pm 5)$ nm. The shrinkage is 1.1 for both, $a$ and $h$. The mass density of non-porous Co$_3$O$_4$ is $\rho = 6.11$ g/cm$^3$. Therefore, the shrunk volume would in that case contain $5.5*10^{-17}$ mol cobalt, much *more* than it started with. This is impossible and proves the existence of pores. However, there is no evidence for lost material yet. Assuming that none was lost, the density is $\rho_{porous} = 4.1$ g/cm$^3$. The specific pore volume *v* can now be calculated from $\rho_{porous} = [v + (1/\rho)]^{-1}$. Note that the stacking of plates leads to pores in between plates. Barrett-Joyner-Halenda (BJH) measurements therefore lead to average pore diameters that are far too large (for example 8 nm, although heights *h* are only ~ 19 nm).

Calcination at 300 to 350 °C shrinks *a* and *h* to 111 and 19 nm. The shrinkage of *h* is 1.4, *higher* than the 1.2 for *a*. This is consistent with a missing surface layer of about 2 nm, but it is still not conclusive evidence. Height shrinks more than other directions if pores migrate, because a randomly migrating pore is likely to reach one of the two faces of the nano plate, while the edges are on average ten times further away than the faces. Assuming the density of solid $Co_3O_4$ again, the amount of Co would be $3.9*10^{-17}$ mol. This equals the initial amount, but the shapes are very porous. In other words, the initial amount was obtained although the pore volume was not yet subtracted. This is clear evidence for missing material. Yet differently put: Under the assumption that no material was lost, the density would be 5.7 g/cm$^3$, much larger than the 4.1 g/cm$^3$ above. Seen this way, one has two conclusions simultaneously: There is material lost now (H1) and the plates after treatment at 200 to 250 °C are internally at least as porous as they are after 300 to 350 °C. Nevertheless, it is well known that BET and electrochemical measurements show 300 °C to be superior. This confirms that the explosions mainly bring about *access* to the pore network and that this access is most important for the desired properties.

## Conclusion

Two predictions from a hypothesized explosive pore formation mechanism were confirmed by conclusive evidence for the lost material (H1) and for the predicted dependence of the accessible specific surface area on the temperature ramping rate (H2). The known optimum pore structure at 300°C has therefore been fully understood, because it is now known why it is worse also at temperatures that are lower than the optimum. It was proved quantitatively that the explosions provide a more accessible pore network, not more pores. The temperature ramping rate dependence can be exploited to provide a material with a far superior specific surface area (60% above previous reports according to initial, not yet optimized measurements). Computer facilitated image analysis was crucial. The shapes are truncated triangles, not regular hexagons. The method exploits information contained in the size distributions' shapes and widths, reducing errors from above 20% down to only 2%. The general method obtained yet again otherwise not attainable quantities such as the weight of single nano shapes and thus the number of shapes, here $4.5*10^{11}$ per mg.

**Supporting Information for publication:** Details of the computationally facilitated image analysis are available in the Electronic Supplementary Information. (see below in this archive version)

## Supplemental Material

**Computer Facilitated Analyses:** The analysis algorithms are programmed with Mathematica9. We color mark the SEM images with the line draw tool of the "Paint" accessory program. For example, the scale bar is turned green (Fig. 2c). The image analysis algorithm first extracts the green components into a new black and white image called "ImageG". The code "ScalebarL = ComponentMeasurements[ImageG, "Length"][[1,2]]" will therefore assign the variable "ScalebarL," which is some number that is proportional to the scale bar's length. Similar short bits of code extract blue colored lines and so on. The algorithm multiplies with the known length of the scale bar and divides by ScalebarL. It therefore knows all lengths in the image, or more precisely: the lengths' projections onto the x-y plane. Lengths that are not flat inside the x-y plane appear shorter than they really are. In order to calculate the physical lengths, one must know a shape's tilt relative to the image pane. The shapes obstruct each other from view (Fig. 2a). Only those shapes that allow marking the necessary consecutive sides (edges) are marked. The edges are marked by blue lines. Marking two sequential edges provides two lengths and one enclosed angle at the branching point between them, together three degrees of freedom (DoF). Graph-analysis (Fig. 2d) is enabled by pointing out the enclosed branching points (nodes, vertices) with short yellow lines. Heights $h$ are separately analyzed; we mark heights of shapes that are clearly seen edge on. With regular hexagons, three DoF suffice, because the single edge length $a$ is only one DoF and 3D orientation is determined by two Euler angles. Such an analysis fails as discussed and the products must be described as truncated triangles (TT).

Consider a TT that lies flat on the x-y-plane (red outline in Fig. 2a). Edge *a* is truly longer than the two adjacent edges *b*. A regular hexagon that is rotated out of view by a large angle around one of its edges *a* has the two neighbouring edges being partially going along the z-direction orthogonal to the image, thus the adjacent sides do also appear shorter (black outline in Fig. 2a). Distinguishing these two shapes needs at least four DoF (two Euler angles and *a* and *b*). We therefore mark three consecutive edges, which together with the two enclosed angles provide five DoF. Theoretically, we now only require solving three vector relations. One is the dot-product between the first and the third edge, call them ***a*** [the vector ***a*** = $(a_1, a_2, a_3)$ with length $a = (\mathbf{a}.\mathbf{a})^{1/2}$] and ***c***. It equals $\mathbf{a}.\mathbf{c} = -a^2/2$, because $c = a$, and the angles between adjacent edges are 60°. Obviously, $(\mathbf{a} + \mathbf{c})/a = \mathbf{b}/b$, which suggests to calculate the ratio $b/a$ via the vector components, namely $b/a = b_1/(a_1 + c_1)$ [Eq. (S1)]. Testing the algorithm revealed that whenever $c_1 = -a_1$ almost holds, Eq. (S1) puts $b/a \approx b_1/0$. This overwhelms the statistics with large values for *b* which are due to those shapes that happen to have $c_1$ very close to $-a_1$ while $b_1$ is still not zero. Avoiding this requires all five DoF. We calculate $a^2$ via a quadratic equation, and $(b/a)^2$ via $(b_1^2 + b_2^2)/[(a_1 + c_1)^2 + (a_2 + c_2)^2]$ to prevent division by small numbers. In order to ensure that the chosen one of the two possible quadratic solutions is correct, that the zero-division problem is solved correctly, and that no other issue was overlooked, we tested the algorithm again with simulated images (Fig. 2b). The finite resolution becomes only problematic at tilt angles over about 85°. Such obviously strongly tilted shapes are anyway not marked for the extraction of lengths *a* and *b*. They instead enter the calculation of the height *h*.

We first analyzed three different SEM per sample, every SEM having up to 32 shapes' edges marked. The resulting three averages never deviate from the combined average by more than the smallest of the three standard deviations. The statistical ensemble of the combined images is therefore sufficiently large. The deviations between different SEM of the same sample are larger than the errors due to the shortcomings of the image analysis method (as applied to single images, where one could count any deviation as a systematic error, since the output will be the same if re-analyzing the same image). E.g.: the random error due to marking a few pixels less or more is smaller than these deviations. The reported statistical error therefore reflects the statistical variation of the physical sizes.